\documentclass[aps,pra,twocolumn,groupedaddress,amsmath,amssymb]{revtex4}
\usepackage{graphicx}
\usepackage{tensor}
\usepackage{xcolor}
\usepackage{mathtools}
\usepackage{mathdots}
\usepackage{bm}

\begin{document}

\title{Statistics of photon-subtracted and photon-added states}

\author{Stephen M. Barnett$^1$, Gergely Ferenczi$^{1,2}$, Claire R. Gilson$^3$ and Fiona C. Speirits$^1$}

\address{$^1$School of Physics and Astronomy, University of Glasgow, Glasgow G12 8QQ, UK}
\address{$^2$Department of Physics, University of Strathclyde, Glasgow G4 0NG, UK}
\address{$^3$School of Mathematics and Statistics, University of Glasgow, Glasgow G12 8SQ, UK}

\date{\today}

Physical Review A - {\it in press}

\begin{abstract}

The subtraction or addition of a prescribed number of photons to a field mode
does not, in general, simply shift the probability distribution by the number
of subtracted or added photons.  Subtraction of a photon from an initial
coherent state, for example, leaves the photon statistics unchanged and the
same process applied to an initial thermal state \emph{increases} the mean
photon number. We present a detailed analysis of the effects of the initial
photon statistics on those of the state from which the photons have been
subtracted or to which they have been added.  Our approach is based on two
closely-related moment generating functions, one well-established and one
that we introduce.

\end{abstract}

\maketitle

\section{Introduction}

The addition or subtraction of a single photon from the radiation field is the
most fundamental process by which matter interacts with light.  The ability to
achieve this level of control in experiments has been used to produce novel
non-classical states of light by the process of ``degaussification"
\cite{Philippe} and in a direct demonstration of the commutation relation
between the annihilation and creation operators
\cite{Bellini1,Bellini2,Bellini3}.

There has been considerable interest in both the processes of photon addition
and subtraction and also in the properties of the quantum states produced by
these processes.  Indeed, a discussion of these appears in Agarwal's textbook
\cite{Girishbook}. Four developments make these states worthy of further
consideration. First there is the rapid advance towards practical
implementation of quantum key distribution~\cite{Wootters,Assche,QIbook} and
the associated eavesdropping activities including photon removal.  Second is
the demonstration, recently, of the effects of photon subtraction on the
visibility of optical fringes with thermal light~\cite{Bobexpt,Lanz} and, more
generally, the suggestion that both photon-subtracted and photon-added states
may provide advantages in metrology~\cite{Treps}. Third
is the requirement for non-Gaussian processes (including photon subtraction or
addition) in order to demonstrate the supremacy of continuous variable quantum
computing~\cite{Eisert,Loock,Douce,Walschaers}. Finally, and perhaps most intriguing, is
the demonstration that photon subtraction from a thermal pulse results in an
increase in the energy and that this information can be used for the extraction
of work~\cite{Demon}.

There have been a number of earlier studies of photon-added, and particularly
of photon-subtracted states. Interest in these states appears to originate with
the work of Agarwal and Tara~\cite{Tara,AgarwalNeg}. Implementing this
technique has been shown to introduce novel quantum effects including the
generation of novel superposition states
\cite{Jeong,Ourjoumtsev,Neergaard,Biswas}.  More recently, attention has turned
to the results of multiple addition and subtraction events and how these might
be used to engineer the properties of the light~\cite{Fiurasek,Bogdanov}.

We note that uncontrolled photon subtraction events arise in the quantum jumps
approach to dissipation~\cite{Carmichael1,Dalibard,Gardiner,Carmichael2} and
these have been shown to have a dramatic effect, in particular, on
non-classical phenomena including Schr\"{o}dinger-cat states~\cite{Garraway}
and on revivals in the Jaynes-Cummings model~\cite{JCM1,JCM2}.  Here, however,
our focus will be on processes designed to subtract or add a given number of
photons even though the probability for the process to be successful will
typically be small.

In this paper we aim to give a thorough description of the photon statistics of
photon-added and photon-subtracted states.  Our preferred tools for this are
the moment generating function, as advocated by Bogdanov \textit{et al}
\cite{Bogdanov}, and a closely related function which we introduce. We find
that a combination of these allows us to make very general statements about the
effects of subtracting or adding a given number of photons and also about the
relationships between these two processes.

\section{Photon-added and photon-subtracted states}

We shall be concerned, for simplicity, solely with states of a single quantised
field mode and the effect of successfully either subtracting or adding one or
more photons to the state of the field. We denote by $\hat\rho^{0}$ the initial
state of the field mode and then the subtraction or addition of a single photon
will produce a state with density operator
\begin{eqnarray}
\label{Eq1.1}
\hat\rho^{1-} &=& \frac{\hat a\hat\rho^{0}\hat a^\dagger}{{\rm Tr}(\hat\rho^{0}\hat a^\dagger \hat a)}  \nonumber \\
\hat\rho^{1+} &=& \frac{\hat a^\dagger\hat\rho^{0}\hat a}{{\rm Tr}(\hat \rho^{0}\hat a \hat a^\dagger)} ,
\end{eqnarray}
respectively.  Adding or subtracting more than a single photon in this way is
challenging, experimentally, but it is, nevertheless, interesting to consider
this possibility at least theoretically, principally for the insights into the
nature of the statistics, to consider states in which more than a single photon
is subtracted or added.  We denote the states following the subtraction or
addition of $\ell$ photons as
\begin{eqnarray}
\label{Eq1.2}
\hat\rho^{\ell-} &=& \frac{\hat a^\ell\hat\rho^{0}\hat a^{\dagger \ell}}{{\rm Tr}(\hat\rho^{0}\hat a^{\dagger \ell} \hat a^\ell)}  \nonumber \\
\hat\rho^{\ell+} &=& \frac{\hat a^{\dagger \ell}\hat\rho^{0}\hat a^\ell}{{\rm Tr}(\hat\rho^{0}\hat a^\ell \hat a^{\dagger \ell})} .
\end{eqnarray}
That these photon-added states, in particular, are worthy of further
consideration was suggested many years ago by Agarwal and Tara~\cite{Tara}.

\begin{figure}
\centering
\includegraphics{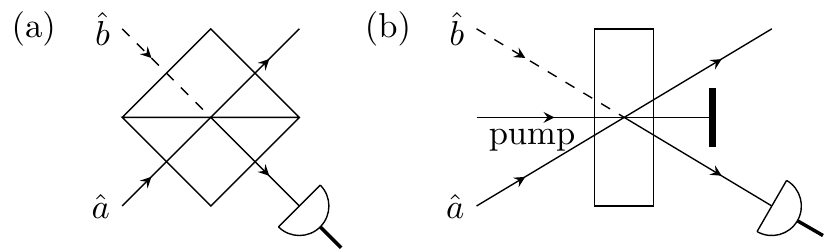}
\caption{Schematic for implementations of (a) photon subtraction using a weakly
reflecting beam splitter and (b) photon addition using a weak parametric amplifier.}
\label{fig:figure1}
\end{figure}

We note that producing photon-subtracted or photon-added states of the form
under consideration is necessarily a probabilistic process with, typically, a
low probability of success.  There are processes that remove a photon with
certainty, if at least one photon is present~\cite{Calsamiglia,Oi,Rosenblum},
but these are more difficult to implement than the probabilistic processes and
will not concern us here.  The simplest way to either subtract or add a single
photon is by a weak interaction with an ancillary mode, with the detection of a
photon in this additional mode heralding a successful subtraction or addition
event~\cite{Girishbook}.  For completeness, we summarise briefly these two
processes.  To realise photon subtraction we combine our mode, $\hat{a}$, with
a second one, $\hat{b}$, prepared in its vacuum state, using a
weakly-reflecting beam splitter as depicted in Fig.~\ref{fig:figure1}(a).  We
can describe the action of the beam splitter by a unitary transformation
coupling the two modes~\cite{Methods}:
\begin{equation}
\label{Eq1.a}
\hat{U} = \exp\left[i\theta(\hat{a}^\dagger \hat{b} + \hat{b}^\dagger \hat{a})\right] .
\end{equation}
The action of this on the two input modes produces the state
\begin{eqnarray}
\label{Eq1.b}
\hat{U}\hat{\rho}^0\otimes|0\rangle\langle 0|\hat{U}^\dagger \approx  (1 + i\theta \hat{b}^\dagger \hat{a})
\hat{\rho}^0\otimes|0\rangle\langle 0|(1 - i\theta \hat{a}^\dagger \hat{b}) .
\end{eqnarray}
If we detect a photon in the output $b$ mode then the output $a$ mode
conditioned on this detection will be the photon-subtracted state
$\hat{\rho}^{1-}$.  To realise photon addition we proceed in the same way but
utilise a weak nonlinear optical parametric-amplfication process, as depicted
in Fig.~\ref{fig:figure1}(b), rather than a beam splitter.  We can describe
this process by a unitary transformation of the form~\cite{Methods}
\begin{equation}
\label{Eq1.c}
\hat{U} = \exp\left[i\vartheta(\hat{a}^\dagger \hat{b}^\dagger + \hat{b}\hat{a})\right] .
\end{equation}
The action of this on the two input modes, with mode $b$ again prepared in the
vacuum state, produces the two-mode output state
\begin{eqnarray}
\label{Eq1.d}
\hat{U}\hat{\rho}^0\otimes|0\rangle\langle 0|\hat{U}^\dagger \approx  (1 + i\vartheta \hat{a}^\dagger \hat{b}^\dagger)
\hat{\rho}^0\otimes|0\rangle\langle 0|(1 - i\vartheta \hat{b} \hat{a}) .
\end{eqnarray}
As with the photon-subtraction process, if we detect a photon in the output $b$
mode then the output $a$ mode conditioned on this detection will be the
photon-added state $\hat{\rho}^{1+}$.  We can produce, at least in principle,
multiple photon-subtracted or photon-added states by combining a number of
single-photon subtraction or addition events accepting, of course, the fact
that the probability for successfully adding or subtracting the photons falls
off rapidly as the number of subtraction or addition events increases.

%%%%%%%%%%%%%%%%%%%%%%%%%%%%%%%%%%%%%%%%%%%%%%%%%%%

%%%%%%%%%%%%%%%%%%%%%%%%%%%%%%%%%%%%%%%%%%%%%%%%%%%

We present in this paper a detailed study of the statistics of
photon-subtracted and photon-added states and provide a simple and efficient
way of obtaining these by expressing the properties of these states in terms of
those of the preprocessed state $\hat\rho^{0}$.  As a foretaste of this we
prove two simple properties.  The first of these is the well-known fact that
subtracting a single photon can result in an increase in the mean photon number
\cite{Ueda,Ban,Zavatta,Dodonov} 
and there is a simple and general criterion for this to occur.
The second is that adding a single photon results in the mean photon number
increasing by at least one. The mean photon number for the photon-subtracted
state is~\cite{Bogdanov}
\begin{equation}
\label{Eq1.3}
\langle \hat n \rangle^{1-} = \frac{{\rm Tr}(\hat\rho^0 \hat a^{\dagger 2}\hat a^2)}{{\rm Tr}(\hat\rho^0 \hat a^\dagger \hat a)} .
\end{equation}
This will be greater than the mean photon number for the original state, ${\rm
Tr}(\hat\rho^0 \hat a^\dagger \hat a)$, if the second-order coherence function
for $\hat\rho^0$
\begin{equation}
\label{Eq1.4}
g^{(2)} = \frac{{\rm Tr}(\hat\rho^0 \hat a^{\dagger 2}\hat a^2)}{[{\rm Tr}(\hat\rho^0 \hat a^\dagger \hat a)]^2}
\end{equation}
is greater than unity, corresponding to a super-Poissonian state, one with a
photon-number variance exceeding the mean value: $\Delta n^2 > \langle \hat n
\rangle$~\cite{Demon,Ueda,Zavatta}.  For the photon-added state the mean value of the photon
number is
\begin{eqnarray}
\label{Eq1.5}
\langle \hat n \rangle^{1+} &=& \frac{{\rm Tr}(\hat\rho^0 \hat a \hat a^{\dagger}\hat a \hat a^\dagger)}
{{\rm Tr}(\hat\rho^0 \hat a \hat a^\dagger)}
\nonumber \\
&=& \frac{{\rm Tr}[\hat\rho^0 (\hat n+1)^2]}{{\rm Tr}[\hat\rho^0 (\hat n+1)]}  .
\end{eqnarray}
As the mean square of a quantity (in this case $\hat n+1$) is greater than or
equal to the square of the mean it necessarily follows that photon addition
will increase the mean photon number by at least one: $\langle \hat n
\rangle^{1+} \geq \langle \hat n \rangle^{0} + 1$.

\section{Moment generating functions}

Our aim is to determine, in a general manner and as simply as possible, the
relationship between the photon statistics of the initial state, with density
operator $\hat\rho^0$, and that of a state following the subtraction or
addition of a given number of photons.  The most natural tool to use for this
is a moment generating function, as is so often the case in statistics
\cite{Jeffreys,Lukacs,Helstrom,Renyi,vanKampen,GardinerStoch,Grimmett}. We
shall employ a pair of closely related moment generating functions:
\begin{eqnarray}
\label{Eq2.1}
\mathcal{M}(\mu) &=& \sum_{n=0}^\infty (1-\mu)^n P(n)  \nonumber  \\
\mathcal{N}(\lambda) &=& \sum_{n=0}^\infty (1+\lambda)^{-(n+1)} P(n) ,
\end{eqnarray}
where $P(n)$ is the probabililty that $n$ photons are present.  The first of
these is the familiar quantity introduced into quantum optics by Glauber (and
denoted by him as $Q(s)$)~\cite{Glauber,JakemanPike}. The second, although
clearly simply related to the first is, to the best of our knowledge, new to
quantum optics and is introduced here because of its use in treating
photon-added states. Some of the properties of these functions are reviewed in
Appendix \ref{AppendixA}.  The main properties of these that we shall exploit
are that $\mathcal{M}(\mu)$ and $\mathcal{N}(\lambda)$ give, resepctively, the
factorial moments, $\langle \hat n^{(m)}\rangle = \langle \hat{n}(\hat n - 1)
\cdots (\hat n - m + 1)\rangle$, and the negative factorial moments, $\langle
(\hat n + 1)^{(-m)}\rangle = \langle (\hat{n} + 1)(\hat n +2) \cdots (\hat n +
m)\rangle$, simply by differentiation:
\begin{eqnarray}
\label{Eq2.2}
\langle \hat n^{(m)} \rangle &=& \left. \left(-\frac{d}{d\mu} \right)^m \mathcal{M}(\mu)\right|_{\mu = 0}  \nonumber \\
\langle (\hat n + 1)^{(-m)}\rangle &=& \left. \left(-\frac{d}{d\lambda} \right)^m \mathcal{N}(\lambda)\right|_{\lambda = 0} ,
\end{eqnarray}
and that both functions give, straightforwardly, the probability that the
photon number is even or odd~\cite{Methods}:
\begin{equation}
\label{Eq2.3}
\mathcal{M}(2) = P({\rm even}) - P({\rm odd}) = -\mathcal{N}(-2) .
\end{equation}

An important feature of the moment generating functions is the comparative ease
with which we can find these quantities for important quantum states of light.
We illustrate this point by presenting these for five commonly used types of
state: the number (or Fock) states, the coherent states, the thermal or chaotic
states, the squeezed vacuum states and finally the Schr\"{o}dinger cat states.  
For the photon-number state
$|N\rangle$ the probability distribution is simply $P(n) = \delta_{n,N}$ and we
have
\begin{eqnarray}
\label{Eq2.4}
\mathcal{M}_{|N\rangle}(\mu) &=& (1-\mu)^N  \nonumber \\
\mathcal{N}_{|N\rangle}(\lambda) &=& (1 + \lambda)^{-(N+1)} .
\end{eqnarray}
The coherent states, $|\alpha\rangle$, have a Poissonian photon number
probability distribution~\cite{Methods,Rodney}, $P(n) =
e^{-|\alpha|^2}|\alpha|^{2n}/n!$, and the moment generating functions are
\begin{eqnarray}
\label{Eq2.5}
\mathcal{M}_{|\alpha\rangle}(\mu) &=& e^{-\mu|\alpha|^2}  \nonumber \\
\mathcal{N}_{|\alpha\rangle}(\lambda) &=& \frac{1}{1+\lambda}\exp\left(-\frac{\lambda|\alpha|^2}{1+\lambda}\right) .
\end{eqnarray}
It is straightforward to use these to calculate the factorial moments from
the moment generating functions using (\ref{Eq2.2}).  For the
positive moments we find the familiar form \cite{Methods,Rodney}
\begin{equation}
\label{Eq2.5a}
\langle \hat{n}^{(m)}\rangle_{|\alpha\rangle} =  |\alpha|^{2m} .
\end{equation}
The negative moments have a more complicated form and we give, here,
only the first two of these:
\begin{eqnarray}
\label{Eq2.5b}
\langle (\hat{n}+1)^{(-1)}\rangle_{|\alpha\rangle} &=& |\alpha|^2 + 1  \nonumber \\
\langle (\hat{n}+1)^{(-2)}\rangle_{|\alpha\rangle} &=& |\alpha|^4 + 4|\alpha|^2 + 2 .
\end{eqnarray}
Higher order moments are readily obtained by further differentiation of 
$\mathcal{N}_{|\alpha\rangle}(\lambda)$.  The moment generating function
shows, also, that all coherent states have a greater probability that the photon 
number is even than that it is odd as $\mathcal{M}(2) = -\mathcal{N}(-2)
= e^{-2|\alpha|^2}$.

The thermal state is mixed with a density operator that is diagonal in the
number-state basis.  The probability that there are $n$ photons has the
familiar Bose-Einstein form, $P(n) = \bar{n}^n/(\bar{n}+1)^{n+1}$, where
$\bar{n}$ is the mean photon number.  For this state the moment generating
functions are
\begin{eqnarray}
\label{Eq2.6}
\mathcal{M}_{\rm th}(\mu) &=& \frac{1}{1+\mu\bar{n}}  \nonumber \\
\mathcal{N}_{\rm th}(\lambda) &=& \frac{1}{1 + \lambda(\bar{n} + 1)} .
\end{eqnarray}
The positive and negative moments for the thermal state, derived from these
moment generating functions, have the simple forms:
\begin{eqnarray}
\label{Eq2.6a}
\langle \hat{n}^{(m)}\rangle_{\rm th} &=& m!\bar{n}^m \nonumber \\
\langle (\hat{n}+1)^{(-m)}\rangle_{\rm th} &=& m!(\bar{n}+1)^m .
\end{eqnarray}
Like the coherent states, all thermal states have a higher probability that the 
photon number is even than that it is odd: $\mathcal{M}(2) = -\mathcal{N}(-2)
= (1 + 2\bar{n})^{-1}$.

A much-studied and important non-classical state is the squeezed vacuum,
$|\zeta\rangle$, which is a superposition of only even photon-number states
\cite{Methods,Rodney}:
\begin{eqnarray}
\label{Eq2.7}
P_{|\zeta\rangle}(2n) &=& \frac{1}{\sqrt{\bar{n}+1}}\left(\frac{\bar{n}}{\bar{n}+1}\right)^{n}\frac{(2n)!}{2^{2n}(n!)^2} \nonumber \\
P_{|\zeta\rangle}(2n+1) &=& 0 ,
\end{eqnarray}
where $\bar{n}$ is the mean photon number.  For this state the moment
generating functions are
\begin{eqnarray}
\label{Eq2.8}
\mathcal{M}_{|\zeta\rangle}(\mu) &=& (1 + 2\mu\bar{n} - \mu^2\bar{n})^{-1/2}  \nonumber \\
\mathcal{N}_{|\zeta\rangle}(\lambda) &=& \frac{1}{1+\lambda}\left(1 +
\frac{\lambda(2+\lambda)}{(1+\lambda)^2}\bar{n}\right)^{-1/2} .
\end{eqnarray}
Note that for this state $\mathcal{M}(2) = 1 = -\mathcal{N}(-2)$, which
reflects the fact that the photon number is even.  For the squeezed vacuum
state the first two positive and negative factorial moments, calculated
from the moment generating function, are
\begin{eqnarray}
\label{Eq2.8a}
\langle\hat{n}^{(1)}\rangle &=& \bar{n}  \nonumber \\
\langle\hat{n}^{(2)}\rangle &=& 3\bar{n}^2 + \bar{n}  \nonumber \\
\langle(\hat{n}+1)^{(-1)}\rangle &=& \bar{n} + 1 \nonumber \\
\langle(\hat{n}+1)^{(-2)}\rangle &=& 3\bar{n}^2 + 5\bar{n} + 2 .
\end{eqnarray}

Our final example is the
even and odd Schr\"{o}dinger cat states, $|\alpha\pm\rangle$, which are
superpositions of a pair of coherent states~\cite{Girishbook}:
\begin{equation}
\label{Eq2.9}
|\alpha\pm\rangle = \frac{1}{\sqrt{2(1\pm e^{-2|\alpha|^2})}} \left(|\alpha\rangle \pm |-\alpha\rangle\right) .
\end{equation}
Among the interesting properties of these states is the fact that they are
superpositions of only even photon numbers
\begin{eqnarray}
\label{Eq2.10}
P_{|\alpha+\rangle}(2n) &=& \frac{1}{\cosh |\alpha|^2}\frac{|\alpha|^{4n}}{(2n)!}  \nonumber \\
P_{|\alpha+\rangle}(2n+1) &=& 0 \, ,
\end{eqnarray}
or odd photon numbers respectively
\begin{eqnarray}
P_{|\alpha-\rangle}(2n) &=&  0   \nonumber \\
P_{|\alpha-\rangle}(2n+1) &=& \frac{1}{\sinh |\alpha|^2}\frac{|\alpha|^{2(2n+1)}}{(2n+1)!} .
\end{eqnarray}
It is straightforward to calculate the forms of our two moment generating
functions for these states. For the even Schr\"{o}dinger cat state we find
\begin{eqnarray}
\label{Eq2.11}
\mathcal{M}_{|\alpha+\rangle}(\mu) &=& \frac{\cosh[(1-\mu)|\alpha|^2]}{\cosh |\alpha|^2}  \nonumber \\
\mathcal{N}_{|\alpha+\rangle}(\lambda) &=& \frac{\cosh [|\alpha|^2/(1+\lambda)]}{(1+\lambda)\cosh |\alpha|^2} ,
\end{eqnarray}
so that $\mathcal{M}(2) = 1 = -\mathcal{N}(-2)$ because the photon number is
even.  For the odd Schr\"{o}dinger cat state, however, our moment generating
functions are
\begin{eqnarray}
\label{Eq2.12}
\mathcal{M}_{|\alpha-\rangle}(\mu) &=& \frac{\sinh[(1-\mu)|\alpha|^2]}{\sinh |\alpha|^2}  \nonumber \\
\mathcal{N}_{|\alpha-\rangle}(\lambda) &=& \frac{\sinh [|\alpha|^2/(1+\lambda)]}{(1+\lambda)\sinh |\alpha|^2} ,
\end{eqnarray}
for which $\mathcal{M}(2) = -1 = -\mathcal{N}(-2)$, which is a consequence of
the fact that only odd photon numbers are present in the odd cat state.
As with our other examples, it is straightforward to use the moment generating 
functions to obtain the positive and negative factorial moments for these states.
For the even cat states we find
\begin{eqnarray}
\label{Eq2.13}
\langle\hat{n}^{(1)}\rangle_{|\alpha_+\rangle} &=& |\alpha|^2\tanh |\alpha|^2  \nonumber \\
\langle\hat{n}^{(2)}\rangle_{|\alpha_+\rangle} &=& |\alpha|^4  \nonumber \\
\langle(\hat{n}+1)^{(-1)}\rangle_{|\alpha_+\rangle} &=& |\alpha|^2\tanh |\alpha|^2 + 1  \nonumber \\
\langle(\hat{n}+1)^{(-2)}\rangle_{|\alpha_+\rangle} &=& |\alpha|^4 + 4|\alpha|^2\tanh |\alpha|^2 + 2 .
\end{eqnarray}
The first two of these mean that the even cat state is super-Poissonian with $\Delta n^2 > 
\langle\hat{n}\rangle$.  For the odd cat states we have
\begin{eqnarray}
\label{Eq2.14}
\langle\hat{n}^{(1)}\rangle_{|\alpha_+\rangle} &=& |\alpha|^2\coth |\alpha|^2  \nonumber \\
\langle\hat{n}^{(2)}\rangle_{|\alpha_+\rangle} &=& |\alpha|^4  \nonumber \\
\langle(\hat{n}+1)^{(-1)}\rangle_{|\alpha_+\rangle} &=& |\alpha|^2\coth |\alpha|^2 + 1  \nonumber \\
\langle(\hat{n}+1)^{(-2)}\rangle_{|\alpha_+\rangle} &=& |\alpha|^4 + 4|\alpha|^2\coth |\alpha|^2 + 2 .
\end{eqnarray}
We see that, in contrast with the even cat states, the odd states are sub-Poissonian with 
$\Delta n^2 < \langle\hat{n}\rangle$.

\section{Statistics of photon-subtracted states}

The simplest way to appreciate the changes in the statistics of a
photon-subtracted state is through the factorial moments.  The $m^{\rm th}$
factorial moment of the photon number is defined to be
\begin{eqnarray}
\label{Eq4.1}
\langle \hat n^{(m)} \rangle &=& \langle \hat n(\hat n-1) \cdots (\hat n-m+1)\rangle  \nonumber \\
&=& \langle \hat{a}^{\dagger m}\hat{a}^m\rangle .
\end{eqnarray}
When written in this form it is readily apparent that the $m^{\rm th}$
factorial moment for the $\ell$-photon subtracted state is simply related to
the $(m+\ell)^{\rm th}$ factorial moment of the initial, pre photon-subtraction
state:
\begin{eqnarray}
\label{Eq4.2}
\langle \hat n^{(m)} \rangle^{\ell -} &=& \frac{{\rm Tr}(\hat{a}^{\dagger m}\hat{a}^m\hat{a}^{\ell}\hat{\rho}^0\hat{a}^{\dagger \ell})}
{{\rm Tr}(\hat{\rho}^0\hat{a}^{\dagger \ell}\hat{a}^\ell)} \nonumber \\
&=& \frac{\langle \hat n^{(m+\ell)} \rangle^0}{\langle \hat n^{(\ell)} \rangle^0} ,
\end{eqnarray}
which is the ratio of the $(m+\ell)^{\rm th}$ and $\ell^{\rm th}$ factorial
moments for the initial, pre photon-subtraction state.

It is natural and straightforward to express the full photon statistics of the
photon subtracted states in terms of the moment generating function
$\mathcal{M}(\mu)$.  To see this we make use of the expression for the moment
generating function in terms of the factorial moments, Eq.~(\ref{EqA10}), to
write $\mathcal{M}(\mu)$ for the $\ell$-photon subtracted state in the form
\begin{eqnarray}
\label{Eq4.3}
\mathcal{M}^{\ell -}(\mu) &=& \sum_{m=0}^\infty \frac{(-\mu)^m}{m!} \frac{\langle \hat n^{(m+\ell)} \rangle^0}
{\langle \hat n^{(\ell)} \rangle^0}  \nonumber \\
&=& \frac{1}{\langle \hat n^{(\ell)} \rangle^0}\left(-\frac{d}{d\mu}\right)^\ell \mathcal{M}^0(\mu) ,
\end{eqnarray}
so the moment generating function for the $\ell$-photon subtracted state is
simply the $\ell^{\rm th}$ derivative of that for the pre photon-subtracted
state, normalised so that $\mathcal{M}^{\ell -}(0) = 1$~\cite{Bogdanov}.

It remains to demonstrate the utility of the simple photon-subtraction
transformation of the moment generating function, which we do by exploring the
effects on the states considered in the preceding section.  The effect of a
successful $\ell$-photon subtraction on the number state $|N\rangle$ is simply
to reduce the photon number to $N-\ell$ and this is reflected in the
corresponding moment generating function, which takes the form
\begin{equation}
\label{Eq4.4}
\mathcal{M}_{|N\rangle}^{\ell-}(\mu) = (1-\mu)^{N-\ell} ,
\end{equation}
which we recognise as the moment generating function for the photon number
state $|N-\ell\rangle$.

\begin{figure}
\centering
\includegraphics[width=6cm]{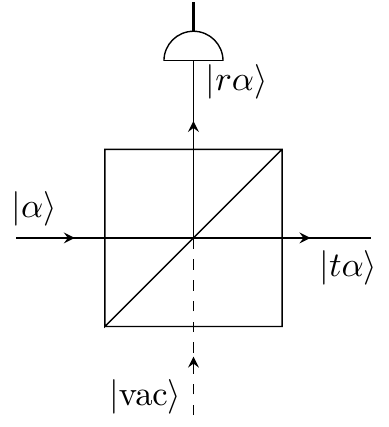}
\caption{The action of our photon subtraction process on an initial coherent state $|\alpha\rangle$.}
\label{fig:figure2}
\end{figure}

The effect of photon subtraction on the coherent state is interesting; the
statistics are unchanged by the process:
\begin{equation}
\label{Eq4.5}
\mathcal{M}_{|\alpha\rangle}^{\ell-}(\mu) = e^{-\mu|\alpha|^2} = \mathcal{M}_{|\alpha\rangle}^{0}(\mu) .
\end{equation}
The reason for this is that the coherent state is a right-eigenstate of the
annihilation operator and hence $(\hat{a})^\ell|\alpha\rangle =
(\alpha)^\ell|\alpha\rangle$, so the $\ell$-photon subtracted coherent state is
simply the initial coherent state.  The physical origin of this unchanging
character under photon subtraction is the fact that a coherent state incident
on a beam-splitter produces two output modes each of which is in a coherent
state, with no entanglement created between the modes.  This process is
depicted in Fig.~\ref{fig:figure2}.  Here an initial coherent state is combined
with a vacuum mode on a very weakly reflecting beam splitter, which enacts the
state transformation $|\alpha\rangle|{\rm vac}\rangle \rightarrow
|t\alpha\rangle |r\alpha\rangle \approx |\alpha\rangle|r\alpha\rangle$.  There
is no correlation between the two output modes and the statistics of the
transmitted mode are independent of whether or not a photocount is recorded at
the detector placed to detect any reflected light.

%%%%%%%%%%%%%%%%%%%%%%%%%%%%%%%%%%%%%%%%%%%%%%%%%%%

%%%%%%%%%%%%%%%%%%%%%%%%%%%%%%%%%%%%%%%%%%%%%%%%%%%

For the thermal states, photon subtraction has a dramatic effect on the
statistics~\cite{Bogdanov} including the increase in the mean photon number
mentioned earlier.  After the successful subtraction of $\ell$ photons, an
initial thermal state with mean photon number $\bar{n}$ will have the moment
generating function
\begin{equation}
\label{Eq4.6}
\mathcal{M}_{\rm th}^{\ell-}(\mu) = (1+\mu\bar{n})^{-(\ell+1)} = [\mathcal{M}_{\rm th}(\mu)]^{\ell+1} .
\end{equation}
This means, in particular, that the mean photon number but also all of the
factorial moments for the photon-subtracted thermal state exceed those for the
initial thermal state:
\begin{eqnarray}
\label{Eq4.7}
\langle \hat n^{(m)} \rangle_{\rm th}^{\ell -} &=& \frac{(m+\ell)!}{\ell !}\bar{n}^m \nonumber \\
&=& \left(\begin{array}{c}
                                        m+\ell\\
                                        \ell
                                        \end{array} \right)    \langle \hat n^{(m)} \rangle_{\rm th}^0 .
\end{eqnarray}
To understand how this happens, we need only note that the photon subtraction
process is more likely to succeed if there are more photons initially present,
hence success in subtracting photons makes it \textit{a posteriori} more likely
that a greater number of photons were present initially.  We return to this
point at the end of this section.

As a final illustration, we turn to the two classes of non-classical state for
which the photon number is either even or odd: the squeezed vacuum and
Schr\"{o}dinger cat states.  For the cat states this is a very simple process -
subtracting an even number of photons leaves the cat-state unchanged, but
taking away an odd number of photons causes an even cat to become an odd cat
and an odd cat is transformed into an even cat state:
\begin{eqnarray}
\label{Eq4.8}
\mathcal{M}_{|\alpha \pm\rangle}^{2\ell -}(\mu) &=& \mathcal{M}_{|\alpha \pm\rangle}(\mu) \nonumber \\
\mathcal{M}_{|\alpha \pm\rangle}^{2\ell +1 -}(\mu) &=& \mathcal{M}_{|\alpha \mp\rangle}(\mu) .
\end{eqnarray}
The same procedure may readily be applied to the squeezed vacuum state, but the
general expression for the moment generating function for the $\ell$-photon
subtracted state is rather unwieldy and we give here expressions only for the
subtraction of one or two photons:
\begin{eqnarray}
\label{Eq4.9}
\mathcal{M}_{|\zeta\rangle}^{1-}(\mu) &=& \frac{1-\mu}{(1+2\mu\bar{n} - \mu^2\bar{n})^{3/2}} \nonumber \\
\mathcal{M}_{|\zeta\rangle}^{2-}(\mu) &=& \frac{1+\bar{n}(3 - 4\mu + 2\mu^2)}
{(1+2\mu\bar{n} - \mu^2\bar{n})^{5/2}(1+3\bar{n})} .
\end{eqnarray}
We note that $\mathcal{M}_{|\zeta\rangle}^{1-}(1) = 0$, corresponding to the
fact that there is no vacuum component in this state and also that
$\mathcal{M}_{|\zeta\rangle}^{1-}(2) = -1$ and
$\mathcal{M}_{|\zeta\rangle}^{2-}(2) = 1$, corresponding to states with only
odd or even photon numbers, as should be the case.

\subsection{Probability of successful photon subtraction}
\label{ProbSucc}

We have seen that the process of photon subtraction has features that might
seem at first to be counter-intuitive. These include the fact that the mean
photon number for an initial thermal state is increased by photon subtraction
and that the Poissonian statistics of a coherent state are unchanged by the
process.  It should be emphasised that these features alone suffice to
demonstrate that the process of photon subtraction, as envisaged here, must be
a probabilistic one, for were it deterministic then photon-number conservation
would, necessarily, produce a reduced number of photons in the output state
\cite{Calsamiglia}.  A simple example may help to clarify this point.  Let us
suppose that we have a mode prepared in a mixture (or a superposition) of
the vacuum and the 10 photon state with equal prior probabilities ($P^0(0)=\frac{1}{2}$,
$P^0(10)=\frac{1}{2}$).  It follows that the initial mean photon number is 5.  If we
succeed in subtracting a photon then the resulting field mode will have a mean
photon number of 9, as the fact that the photon subtraction was successful implies
that there were, on this occasion, 10 photons present initially.

It is interesting to note, however, that the statistics of the
photon-subtracted states allow us to make inferences, using Bayesian reasoning
\cite{Box,Mackay}, concerning the probability of success in the process of
photon subtraction.  To demonstrate this we consider just a single simple
example, the success of photon subtraction for an initial coherent state.  We
know that a successful single-photon subtraction leaves the photon number
probability distribution for an initial coherent state unchanged and hence the
probability that there are $n-1$ photons present given a successful photon
subtraction is
\begin{equation}
\label{Eq4.10}
P^{1-}(n-1|{\rm succ.}) = e^{-|\alpha|^2}\frac{|\alpha|^{2(n-1)}}{(n-1)!} .
\end{equation}
As there was a photon subtraction it follows, necessarily, that this is also
the \textit{posterior} probability that there were $n$ photons present prior to
the subtraction event:
\begin{equation}
\label{Eq4.11}
P^{0}(n|{\rm succ.}) = e^{-|\alpha|^2}\frac{|\alpha|^{2(n-1)}}{(n-1)!} .
\end{equation}
We can use Bayes' theorem to obtain, from this, the probability of successfully
subtracting a photon given that $n$ photons were initially present:
\begin{eqnarray}
\label{Eq4.12}
P^0({\rm succ.}|n)P^0(n) &=& P^0(n|{\rm succ.})P({\rm succ.}) \nonumber \\
\Rightarrow
P^0({\rm succ.}|n)e^{-|\alpha|^2}\frac{|\alpha|^{2n}}{n!} &=& e^{-|\alpha|^2}\frac{|\alpha|^{2(n-1)}}{(n-1)!}P({\rm succ.}) \nonumber \\
\Rightarrow P^0({\rm succ.}|n) &=& n \frac{P({\rm succ.})}{|\alpha|^2} .
\end{eqnarray}
Hence the probability for successfully subtracting a single photon is
proportional to the number of photons initially present. There is a simple
reason for this, which becomes clear on referring to the physical realisation
of the photon subtraction device, based on a weakly reflecting beam splitter:
each photon in the input state is reflected and then detected with a small
probability, $p$, thus the probability that one of the initial $n$ photons
present will be so removed is $np(1-p)^{n-1}$~\cite{Methods}, which, for the
very small reflection probabilities considered here, is approximately $np$.

\section{Statistics of photon-added states}

For the photon-added states it is the negative or ascending factorial moments
\cite{Steffensen},
\begin{eqnarray}
\label{Eq4.13}
\langle (\hat n + 1)^{(-m)}\rangle &=& \langle (\hat n + 1)(\hat n + 2)\cdots (\hat n + m)\rangle \nonumber \\
&=& \langle \hat{a}^m\hat{a}^{\dagger m}\rangle ,
\end{eqnarray}
rather than the more familiar factorial moments that provide the natural
description of the photon statistics. These negative factorial moments are the
expectation values of the antinormal-ordered powers of the number operator
rather than the normally-ordered moments that form the factorial moments. There
is a simple relationship between the negative factorial moments for the
photon-added states and those for the initial state that follows directly from
the form of the states:
\begin{eqnarray}
\label{Eq4.14}
\langle (\hat n + 1)^{(-m)} \rangle^{\ell +} &=& \frac{{\rm Tr}(\hat{a}^{m}\hat{a}^{\dagger m}\hat{a}^{\dagger \ell}\hat{\rho}^0\hat{a}^{ \ell})}
{{\rm Tr}(\hat{\rho}^0\hat{a}^{\dagger \ell}\hat{a}^\ell)} \nonumber \\
&=& \frac{\langle (\hat n + 1)^{(-m-\ell)} \rangle^0}{\langle (\hat n + 1)^{(-\ell)} \rangle^0} ,
\end{eqnarray}
which is the ratio of $(m + \ell)^{\rm th}$ and $\ell^{\rm th}$ negative
factorial moments for the initial, pre photon-addition state.

For the photon-added states it is natural to use our second moment generating
function, $\mathcal{N}(\lambda)$. To construct this quantity we make use of the
expression, Eq.~(\ref{EqA20}), for $\mathcal{N}(\lambda)$ in terms of the
negative factorial moments:
\begin{eqnarray}
\label{Eq4.14a}
\mathcal{N}^{\ell +}(\lambda) &=& \sum_{m=0}^\infty \frac{(-\lambda)^m}{m!}
\frac{\langle (\hat n + 1)^{(-m-\ell)} \rangle^0}{\langle (\hat n + 1)^{(-\ell)} \rangle^0}  \nonumber \\
&=& \frac{1}{\langle (\hat n + 1)^{(-\ell)} \rangle^0}\left(-\frac{d}{d\lambda}\right)^\ell \mathcal{N}^0(\lambda) ,
\end{eqnarray}
so the moment generating function (of the second kind) for the $\ell$-photon
added state is simply the $\ell^{\rm th}$ derivative of that for the pre
photon-added state, normalised so that $\mathcal{N}^{\ell +}(0) = 1$.

We have seen that the process of adding a single photon increases the mean
photon number by at least one. We can also arrive at this as a result of the
general expression Eq.~(\ref{Eq4.14}), by noting that
\begin{eqnarray}
\label{Eq4.15}
\langle (\hat n + 1)^{(-1)} \rangle^{\ell +} &=& \langle (\hat n + 1) \rangle^{\ell +} \nonumber \\
&=& \frac{\langle (\hat{n}+1)\cdots (\hat{n} + \ell + 1)\rangle^0}{\langle (\hat{n}+1)\cdots (\hat{n} + \ell )\rangle^0} \nonumber \\
&=& \ell + \frac{\langle (\hat{n}+1)(\hat{n}+1)\cdots (\hat{n} + \ell)\rangle^0}{\langle (\hat{n}+1)\cdots (\hat{n} + \ell )\rangle^0}
\nonumber \\
&\geq& \ell + \langle(\hat{n} + 1)\rangle^0 ,
\end{eqnarray}
where we have used the inequality derived in Appendix \ref{AppendixB}.  Clearly
$\ell$ photon addition events increase the mean photon number (and indeed the
mean of $\hat{n} + 1$) by at least $\ell$. The equality holds only for an
initial photon number state for which, naturally enough, $\ell$ photon addition
events add precisely $\ell$ photons.  This is reflected in the form of the
moment generating function for the $\ell$ photon added state:
\begin{equation}
\label{Eq4.16}
\mathcal{N}_{|N\rangle}^{\ell +}(\lambda) = (1+\lambda)^{-(N + \ell + 1)} ,
\end{equation}
which is the form for the photon number state $|N + \ell\rangle$.

For all states other than the photon number states,  $\ell$ photon addition
events will increase the mean photon number by more than $\ell$.  The reason
for this can be traced to the fact that the probability for adding a single
photon given that $n$ are initially present is proportional to $n+1$, a feature
that is reflected in the form of the creation operator and has its origins in
Bose symmetry~\cite{Diracbook}.  It follows that the probability that $n+1$
photons are present given a successful photon-addition event is
\begin{equation}
\label{Eq4.17}
P^{1+}(n+1|{\rm succ.}) = \frac{(n+1)P^0(n)}{\langle \hat{n}\rangle^0 + 1} ,
\end{equation}
where the form of the denominator is determined by the requirement that this probability is normalised.  From this it follows that
the mean photon number in the photon-added state is
\begin{eqnarray}
\label{Eq4.18}
\langle \hat{n}\rangle^{1+} &=& \sum_{n=0}^\infty (n+1) P^{1+}(n+1|{\rm succ.}) \nonumber \\
&=& \langle \hat{n}\rangle^0 + 1 + \frac{(\Delta n^2)^0}{\langle \hat{n}\rangle^0 + 1} ,
\end{eqnarray}
which exceeds $\langle \hat{n}\rangle^0 + 1$, corresponding to an increase of
unity, only if $(\Delta n^2)^0 = 0$, that is if the initial state is a photon
number state.

As an example of the effects of photon addition, we consider the effect of
$\ell$ photon additions to a thermal state. For this state we find that the
moment generating function (of the second kind) is
\begin{equation}
\label{Eq4.19}
\mathcal{N}_{\rm th}^{\ell +}(\lambda) = [1 + \lambda(1 + \bar{n})]^{-(\ell + 1)} = [\mathcal{N}_{\rm th}(\lambda) ]^{\ell + 1} .
\end{equation}
Note the similarity between this expression, for photon addition, and that
found for the first moment generating function for an $\ell$-photon subtracted
thermal state, Eq.~(\ref{Eq4.6}).  From this function we can obtain the full
photon statistics of the $\ell$-photon added state.  We find, in particular,
that the $m^{\rm th}$ negative factorial moment may be evaluated by
differentiation of $\mathcal{N}(\lambda)$ with respect to $\lambda$:
\begin{equation}
\label{Eq4.20}
\langle (\hat{n}+1)^{(-m)}\rangle_{\rm th}^{\ell +} = \frac{(m + \ell)!}{\ell !}(1 + \bar{n})^m .
\end{equation}
For the first negative factorial moment following a single photon addition, for
example, we find
\begin{equation}
\label{Eq4.21}
\langle (\hat{n}+1)\rangle_{\rm th}^{1 +} = 2(\langle \hat{n} \rangle^0 + 1) ,
\end{equation}
in agreement with Eq.~(\ref{Eq4.18}).

We conclude this discussion by examining the effects of photon addition on a
coherent state, with its associated Poissonian probability distribution.
Successful completion of $\ell$ photon additions to an initial coherent state
produces a state with photon statistics completely specified by the moment
generating function $\mathcal{N}_{|\alpha\rangle}^{\ell +}$, which we can write
in the closed form
\begin{eqnarray}
\label{Eq4.22}
\mathcal{N}_{|\alpha\rangle}^{\ell +}(\lambda) &=& \frac{\left(-\frac{d}{d\lambda}\right)^\ell \mathcal{N}_{|\alpha\rangle}(\lambda)}
{\left(-\frac{d}{d\lambda}\right)^\ell \mathcal{N}_{|\alpha\rangle}(0)}  \nonumber \\
&=& \exp\left(-\frac{\lambda|\alpha|^2}{1+\lambda}\right)\frac{L_\ell\left(-\frac{|\alpha|^2}{1+\lambda}\right)}
{(1+\lambda)^{\ell + 1}L_\ell(-|\alpha|^2)}  \nonumber \\
&=& \mathcal{N}_{|\alpha\rangle}(\lambda) \frac{L_\ell\left(-\frac{|\alpha|^2}{1+\lambda}\right)}
{(1+\lambda)^{\ell }L_\ell(-|\alpha|^2)} ,
\end{eqnarray}
where $L_\ell(x)$ is the familiar Laguerre polynomial of order $\ell$.  As a
demonstration of this approach to calculating the statistics, the first
negative factorial moment for the state produced by $\ell$ photon addition
events is
\begin{eqnarray}
\label{Eq4.23}
\langle (\hat{n} + 1)^{(-1)}\rangle_{|\alpha\rangle}^{\ell +} &=& \langle (\hat{n} + 1)\rangle_{|\alpha\rangle}^{\ell+}  \nonumber \\
&=& \left. \left(-\frac{d}{d\lambda}\right)\mathcal{N}_{|\alpha\rangle}^{\ell +}(\lambda)\right|_{\lambda = 0} \nonumber \\
&=& |\alpha|^2 + 2\ell + 1 - \frac{\ell L_{\ell-1}(-|\alpha|^2)}{L_\ell(-|\alpha|^2)} .
\end{eqnarray}
For single-photon addition, this becomes $|\alpha|^2 + 2 + \frac{|\alpha|^2}{1
+ |\alpha|^2}$ in agreement with Eq.~(\ref{Eq4.18}).  More generally, the
successful addition of $\ell$ photons has increased the mean photon number by
$2\ell - \frac{\ell L_{\ell-1}(-|\alpha|^2)}{L_\ell(-|\alpha|^2)}$. This
implies that the initial mean photon number \emph{given that} the subtraction
events were successful is increased from $|\alpha|^2$ to $|\alpha|^2 + \ell -
\frac{\ell L_{\ell-1}(-|\alpha|^2)}{L_\ell(-|\alpha|^2)}$. For small amplitude
coherent states, $|\alpha|^2 \ll 1$, this tends to $|\alpha|^2$, but for higher
values, $|\alpha|^2 \gg 1$, it tends to $|\alpha|^2 + \ell$.  This can be
verified using the Bayesian approach outlined in subsection \ref{ProbSucc}.

\section{Case studies}

It remains to demonstrate the utility of the moment-generating techniques
described above. This we do by presenting results for the subtraction or
addition of photons from coherent and thermal states.  We then address the
effects of the processes of optical attenuation or amplification based on the
properties of binomial~\cite{Stoler} and negative binomial states
\cite{negbinomial}.

\subsection{Coherent states}

The coherent states are right-eigenstates of the annihilation operator and, as
we have seen, this means that the states $\hat{\rho}^{\ell -}$ produced from it
by the subtraction of $\ell$ photons are the same coherent states that we
started with and our photon-subtraction process has no effect on the statistics
of a coherent state.  This is not true for photon addition, which markedly
changes the statistics of the state.

The natural way to derive the photon number probability distribution for a
photon-added coherent state is to use the expression Eq.~(\ref{Eq4.14a}) for
our second moment generating function.  Following this procedure we find for
the one-photon added coherent state, the function
\begin{equation}
\label{Eq5A.1}
\mathcal{N}^{1+}_{|\alpha\rangle}(\lambda) = \frac{e^{-|\alpha|^2}}{(1+\lambda)^2(1+|\alpha|^2)}
\exp\left(\frac{|\alpha|^2}{1+\lambda}\right)\left(1 + \frac{|\alpha|^2}{1+\lambda}\right) ,
\end{equation}
from which we can readily extract the corresponding photon number probability
distribution, either by constructing the power series in $(1+\lambda)^{-1}$ or
using Eq.~(\ref{EqA21}):
\begin{equation}
\label{Eq5A.2}
P^{1+}_{|\alpha\rangle}(n) = \frac{e^{-|\alpha|^2}}{1+|\alpha|^2}\left[\frac{|\alpha|^{2(n-1)}}{(n-1)!} + |\alpha|^2 \frac{|\alpha|^{2(n-2)}}{(n-2)!}\right]
\end{equation}
where factorials of negative numbers are to be understood to take an infinite
value.  This probability distribution is a combination of two shifted
Poissonian distributions, one shifted up by one and the other shifted up by
two. For small amplitude coherent states, the former dominates and the mean
photon number is increased by unity in the process.  For large amplitude
coherent states, however, the latter dominates and the mean photon number is
increased by two, in agreement with the behaviour noted in the preceding
section.

%%%%%%%%%%%%%%%%%%%%%%%%%%%%%%%%%%%%%%%%%%%%%%%%%%%

\begin{figure}
\centering
\includegraphics[width=8cm]{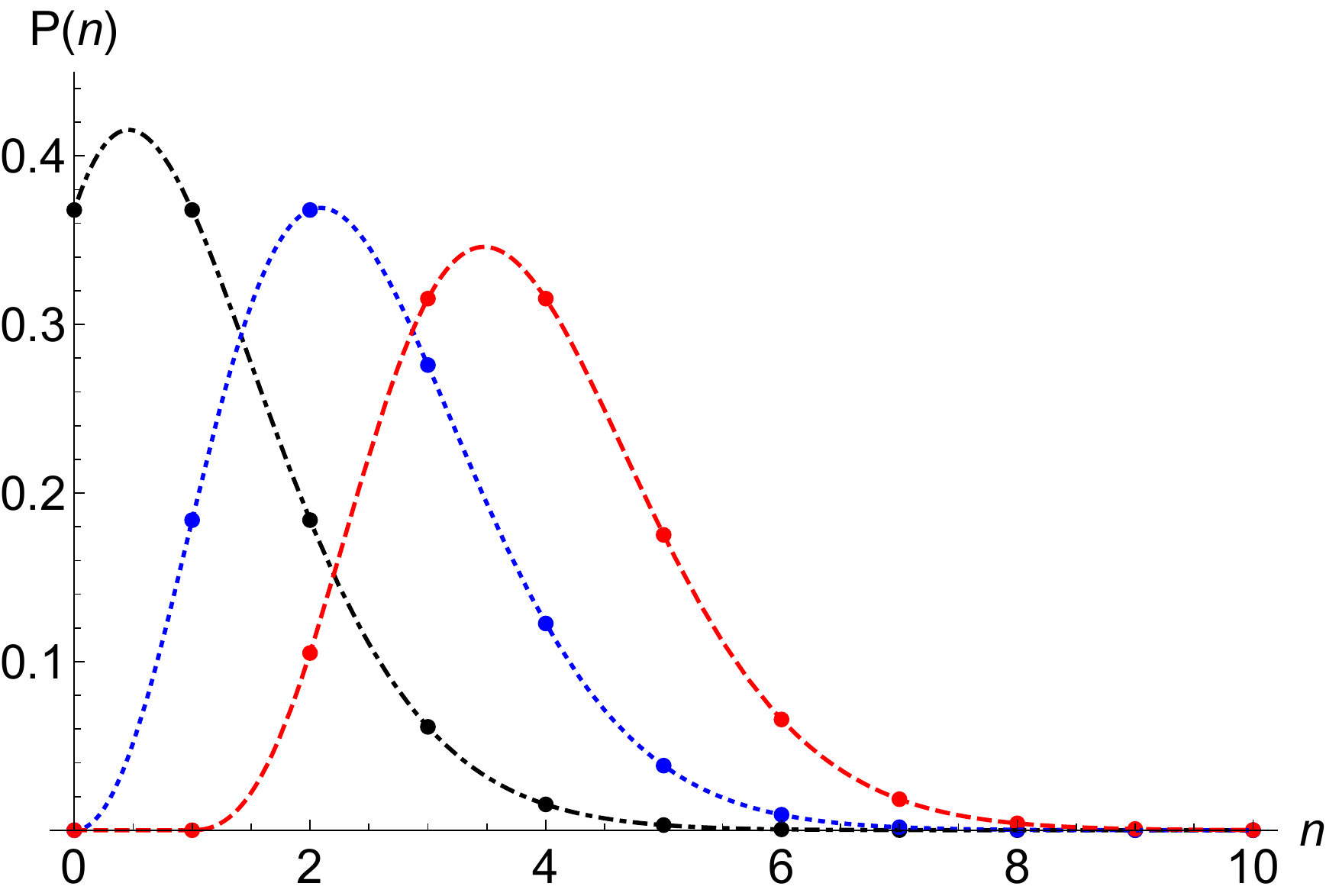}
\caption{(color online)  The photon number probability distributions for (a) an initial coherent state with a
mean photon number of unity (black circles, dash-dotted line) (b) the state produced by a single photon addition
(blue circles, dotted line) and (c) the state produced by two photon addition (red circles, dashed line).  Photon
subtraction leaves the initial coherent-state statistics unchanged.}
\label{fig:figure3}
\end{figure}

%%%%%%%%%%%%%%%%%%%%%%%%%%%%%%%%%%%%%%%%%%%%%%%%%%%

We can extend this technique to find the photon number probability distribution
after any number of photon additions, but present here only the example of two
photon additions.  After two successful photon addition processes our moment
generating function is
\begin{eqnarray}
\label{Eq5A.3}
\mathcal{N}^{2+}_{|\alpha\rangle}(\lambda) &=& \frac{e^{-|\alpha|^2}\exp\left(\frac{|\alpha|^2}{1+\lambda}\right)}
{(1+\lambda)^3(|\alpha|^4+4|\alpha|^2+2)}  \nonumber \\
& & \qquad \times \left(2 + \frac{4|\alpha|^2}{1+\lambda} + \frac{|\alpha|^4}{(1+\lambda)^2}\right) .
\end{eqnarray}
From this we can readily extract the photon number probability distribution:
\begin{eqnarray}
\label{Eq5A.4}
P^{2+}_{|\alpha\rangle}(n) &=& \frac{e^{-|\alpha|^2}}{(|\alpha|^4+4|\alpha|^2+2)}\left[2 \frac{|\alpha|^{2(n-2)}}{(n-2)!} \right.
\nonumber \\
& & \: \: \left. + 4|\alpha|^2 \frac{|\alpha|^{2(n-3)}}{(n-3)!}  + |\alpha|^4\frac{|\alpha|^{2(n-4))}}{(n-4)!} \right] ,
\end{eqnarray}
which comprises three shifted Poissonian distributions, shifted up by 2, 3 and 4 respectively.

In Fig.~\ref{fig:figure3} we plot the photon number probability distributions
for an initial coherent state with a mean photon number of unity and the
distributions that result from successful one- and two-photon addition
processes. The absence of a probability for the vacuum state in the former and
for both the vacuum and one-photon states in the latter is readily apparent. It
is also clear that adding a photon has the effect of broadening the probability
distribution, what may be seen as a consequence of the combination of multiple
shifted Poissonian distributions.

\subsection{Thermal states}

The moment generating functions for the photon-subtracted and photon-added
thermal states have the simple forms given in Eqs.~(\ref{Eq4.6}) and
(\ref{Eq4.19}).  From the similarity in the forms of these it should come as no
surprise that the statistics of an $\ell$-photon subtracted and an
$\ell$-photon added thermal state are simply related.  For this reason it is
sensible to treat them together.

The simplest way to proceed is to expand the two moment generating functions $\mathcal{M}^{\ell-}_{\rm th}(\mu)$ and
$\mathcal{N}^{\ell+}_{\rm th}(\lambda)$ as power series in $1-\mu$ and $1+\lambda$ respectively.  This gives
\begin{eqnarray}
\label{Eq5B.1}
\mathcal{M}^{\ell-}_{\rm th}(\mu) &=& (1+\mu\bar{n})^{-(\ell+1)}  \nonumber \\
&=& \left(\frac{1}{1+\bar{n}}\right)^{\ell+1} \sum_{m=0}^\infty  \left(\frac{(1-\mu)\bar{n}}{1+\bar{n}}\right)^m
\left(\begin{array}{c}
                             \ell + m \\
                             \ell \end{array}\right) , \nonumber \\
& &
\end{eqnarray}
which corresponds to a negative binomial probability
distribution~\cite{Jeffreys,Lukacs,Grimmett} for the photon number
\begin{equation}
\label{Eq5B.2}
P^{\ell-}_{\rm th}(n) = \frac{\bar{n}^n}{(1+\bar{n})^{n+\ell+1}}
\left(\begin{array}{c}
                             \ell + n \\
                             \ell \end{array}\right) .
\end{equation}
For the photon-added thermal states we proceed in the same way but work with $\mathcal{N}(\lambda)$:
\begin{eqnarray}
\label{EqB.3}
\mathcal{N}^{\ell+}_{\rm th}(\lambda) &=& [1 + \lambda(1+\bar{n})]^{-(\ell + 1)}  \nonumber \\
&=& \left(\frac{1}{(1+\lambda)(1+\bar{n})}\right)^{\ell+1}  \nonumber \\
& & \:\:\ \times \sum_{m=0}^\infty \left(\frac{\bar{n}}{(1+\lambda)(1+\bar{n})}\right)^m
\left(\begin{array}{c}
                             \ell + m \\
                             \ell \end{array}\right) ,  \nonumber \\
& &
\end{eqnarray}
corresponding to another negative binomial distribution
\begin{equation}
\label{Eq5B.4}
P^{\ell +}_{\rm th}(n) = \frac{\bar{n}^{(n-\ell)}}{(1+\bar{n})^{n+1}}
\left(\begin{array}{c}
                             n \\
                             \ell \end{array}\right)   \qquad (n \geq \ell).
\end{equation}
For $n<\ell$ the probability is zero, which reflects the fact that $\ell$ photons have been added.

\begin{figure}
\centering
\includegraphics[width=8cm]{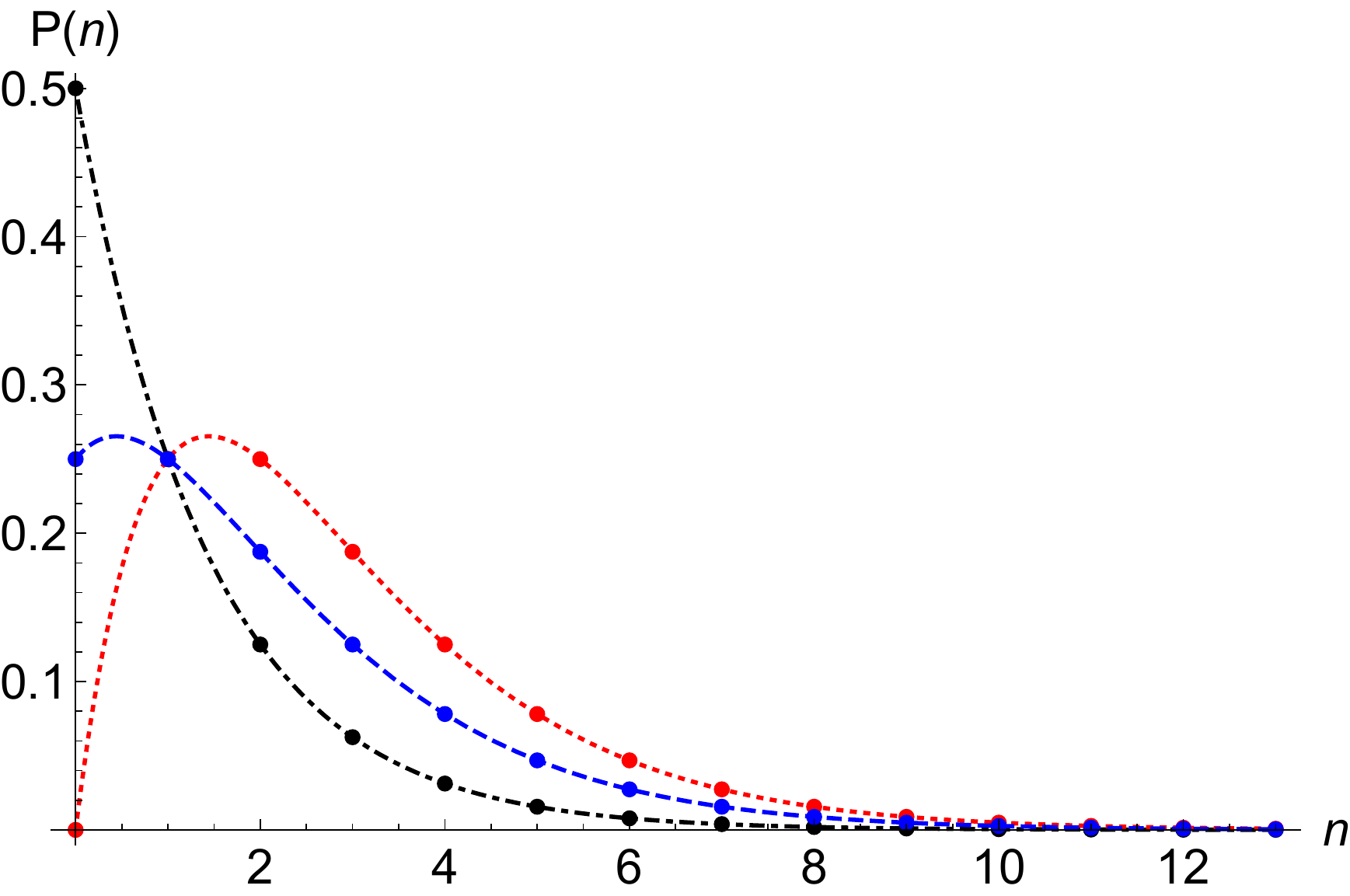}
\caption{(color online)  The photon number probability distributions for (a) an initial thermal state with a
mean photon number of unity (black circles, dash-dotted line) (b) the state produced by a single photon subtraction
(blue circles, dotted line) and (c) the state produced by single photon addition (red circles, dashed line).}
\label{fig:figure4}
\end{figure}

\begin{figure}
\centering
\includegraphics[width=8cm]{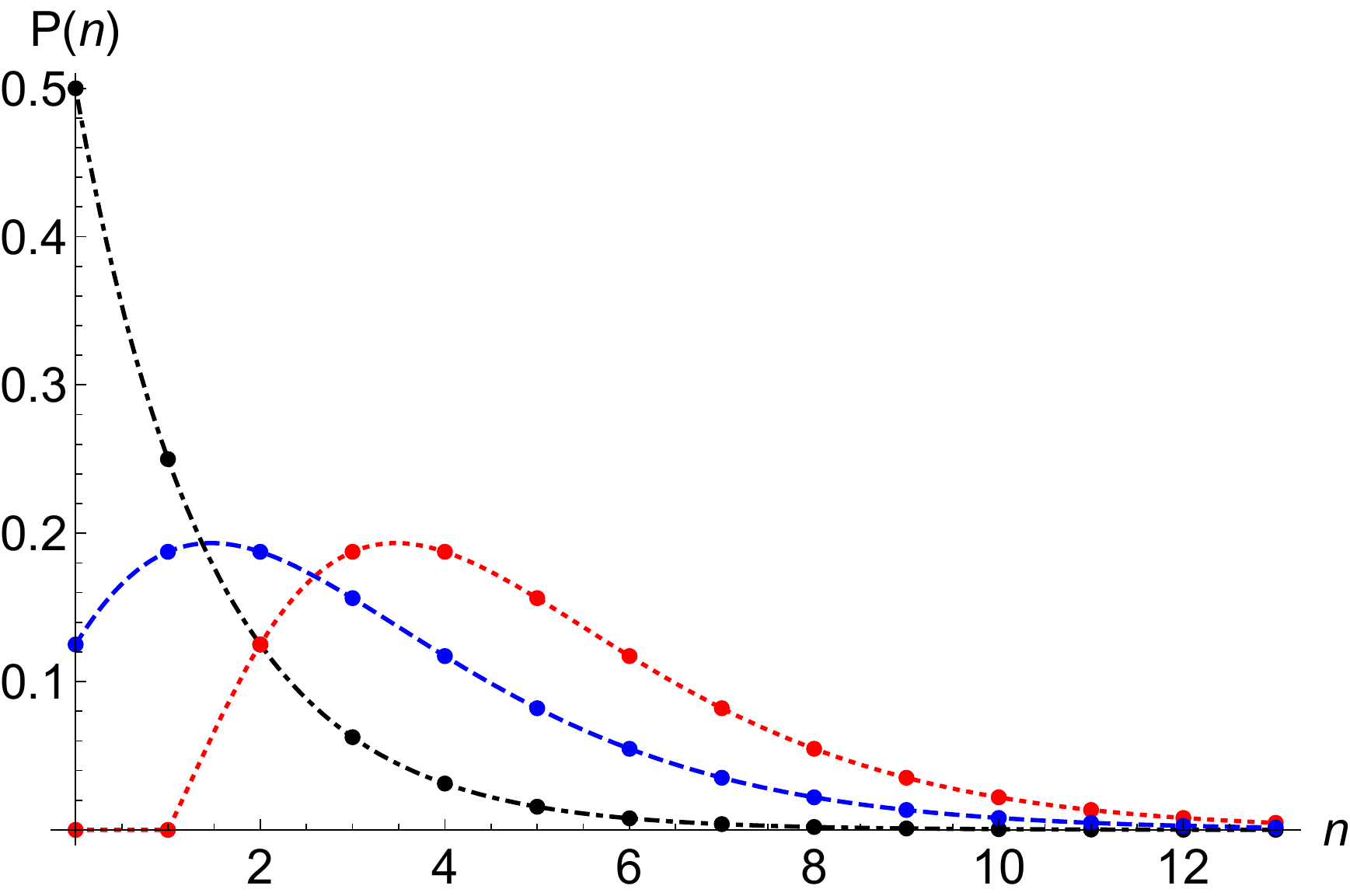}
\caption{(color online)  The photon number probability distributions for (a) an initial thermal state with a
mean photon number of unity (black circles, dash-dotted line) (b) the state produced by a two photon subtraction
(blue circles, dotted line) and (c) the state produced by a two photon addition (red circles, dashed line).}
\label{fig:figure5}
\end{figure}

It is clear that the two photon probability distributions, Eqs.~(\ref{Eq5B.2})
and (\ref{Eq5B.4}) are the same apart from a shift: they have the same shape
but the probability distribution for the photon-subtracted states starts at
zero photons, but that for the photon-added states starts, naturally, at $n =
\ell$.  This behaviour is clear in Figs.~\ref{fig:figure4} and
\ref{fig:figure5}, which show the effects on the statistics of adding or
subtracting one photon and of adding or subtracting two photons, respectively.
The similarity in the distributions means that the statistics of
photon-subtracted and photon-added thermal states are very similar.  In
particular, the mean photon number resulting from $\ell$-photon addition will
exceed that resulting from $\ell$-photon subtraction by precisely $\ell$ and
the variance in the photon number for the two states will be the same.

%%%%%%%%%%%%%%%%%%%%%%%%%%%%%%%%%%%%%%%%%%%%%%%%%%%

%%%%%%%%%%%%%%%%%%%%%%%%%%%%%%%%%%%%%%%%%%%%%%%%%%%

%%%%%%%%%%%%%%%%%%%%%%%%%%%%%%%%%%%%%%%%%%%%%%%%%%%

%%%%%%%%%%%%%%%%%%%%%%%%%%%%%%%%%%%%%%%%%%%%%%%%%%%

\subsection{Binomial and negative binomial states}

Among the most important and most studied quantum optical processes are
attenuation due to propagation through a lossy medium, and amplification using
an inverted population or a parametric amplifier
\cite{Methods,Louisell1,Louisell2,MandelWolf}.  It should be emphasised that
these processes are not simply related to the photon subtraction and addition
processes discussed here.  Rather they are processes formed by random
combinations of successful and unsuccessful subtraction or addition events.

The effect of an ideal (zero-temperature) attenuator is to reduce the factorial
moments by a factor depending on the strength of the attenuation:
\begin{equation}
\label{Eq5.1}
\langle \hat{n}^{(m)}\rangle_{\rm Att.} = \eta^m\langle \hat{n}^{(m)}\rangle ,
\end{equation}
where $0 \leq \eta <1$, with smaller values corresponding to stronger
attenuation.  It follows immediately, on using Eq.~(\ref{EqA10}), that the
moment generating function for the attenuated state has the same form as that
for the pre-attenuated state, but with $\mu$ replaced by $\eta\mu$
\cite{Methods}:
\begin{equation}
\label{Eq5.1a}
\mathcal{M}_{\rm Att.}(\mu) = \mathcal{M}(\eta\mu) .
\end{equation}

Moment generating functions have been used to describe the statistics of
optical amplifiers~\cite{LoudonShep,ShepJakeman}. Here we consider only the
ideal case of a fully-inverted medium amplifier for which the mean photon
number at the output is related to that at the input by
\begin{equation}
\label{Eq5.2}
\langle\hat{n}\rangle_{\rm Amp.} = G\langle\hat{n}\rangle + G-1 ,
\end{equation}
where $G \geq 1$ is the gain.  More generally, we find that negative factorial
moments are related simply to those for the input state
\begin{equation}
\label{Eq5.3}
\langle (\hat{n}+1)^{(-m)}\rangle_{\rm Amp.} = G^m\langle (\hat{n}+1)^{(-m)}\rangle .
\end{equation}
It follows, using Eq.~(\ref{EqA20}), that the moment generating function (of
the second kind) has the same form as that of the pre-amplified state, but with
$\lambda$ replaced by $G\lambda$:
\begin{equation}
\label{Eq5.4}
\mathcal{N}_{\rm Amp.}(\lambda) = \mathcal{N}(G\lambda) .
\end{equation}
The simple expressions, Eqs.~(\ref{Eq5.1a}) and (\ref{Eq5.4}), enable us to
determine the effects of amplification or attenuation on the statistics of our
photon-added states or, indeed, the effects of photon addition or subtraction
on photo-subtracted or photon-added states.  As an illustration, we consider
photon subtraction or addition to an attenuated or amplified photon number
state.  The attenuated number state exhibits binomial statistics and the
amplified number state has negative-binomial statistics.  It is convenient to
investigate these using the binomial~\cite{Stoler} and negative binomial states
\cite{negbinomial}.

\subsubsection{Binomial states}

If we send an $M$-photon state through a lossy medium, in which the probability
for any one photon to survive is $\eta$, then we end up with an incoherent
mixture of number states in which the probability for $n$ photons to remain is
\begin{equation}
\label{Eq5.5}
P^{\rm Att.}_{|M\rangle}(n) =
\left(\begin{array}{c}
                             M \\
                             n \end{array}\right)\eta^n(1-\eta)^{M-n} .
\end{equation}
This mixed state has the same photon statistics as the pure binomial state
$|\eta,M\rangle$~\cite{Stoler}:
\begin{equation}
\label{Eq5.6}
|\eta,M\rangle = \sum_{n=0}^M \left[\left(\begin{array}{c}
                             M \\
                             n \end{array}\right)\eta^n(1-\eta)^{M-n}\right]^{1/2} |n\rangle .
\end{equation}
There is no suggestion that this is the state produced by attenuation, but
merely that it has the same photon statistics. The link with attenuation is
simply one reason to consider the properties of the binomial states.  Some of
the principal properties of the binomial states are summarised in Appendix
\ref{AppendixC}.

The action of the annihilation operator on the binomial state $|\eta,M\rangle$
produces another binomial state, but with $M$ reduced by unity, see
Eq.~(\ref{EqC3}).  It follows that the factorial moments for an $\ell$ photon
subtracted binomial state are simply that for a binomial state with $M$ reduced
by $\ell$:
\begin{eqnarray}
\label{Eq5.7}
\langle \hat{n}^{(m)}\rangle_{|\eta,M\rangle}^{\ell -} &=&  \langle \hat{n}^{(m)}\rangle_{|\eta,M-\ell\rangle}  \nonumber \\
&=& \eta^m \frac{(M-\ell)!}{(M-m-\ell)!} .
\end{eqnarray}
This result has implications for a situation in which both photon addition and
subtraction act to produce the final state. In particular, the form of the
final state does not depend on the whether the subtraction occurs before, after
or during the attenuation. The only difference is the success probability for
the subtraction processes.

\subsubsection{Negative binomial states}

Ideal amplification, with gain $G$, of an initial $M$-photon state produces an
incoherent mixture of number states in which the probability for $n$ photons to
be present is given by the negative binomial distribution,
\begin{equation}
\label{Eq5.8}
P^{\rm Amp.}_{|M\rangle}(n) =
\left(\begin{array}{c}
                             n \\
                             M \end{array}\right)G^{-(M+1)}(1-G^{-1})^{n-M} .
\end{equation}
This mixed state has the same photon statistics as the pure negative binomial
state $|\eta,-(M+1)\rangle$ \cite{negbinomial} with gain $G = \eta^{-1}$
\cite{negbinomial}:
\begin{equation}
\label{Eq5.9}
|\eta, -(M+1)\rangle = \sum_{n=M}^\infty
\left[\left(\begin{array}{c}
                             n \\
                             M \end{array}\right)\eta^{M+1}(1-\eta)^{n-M}\right]^{1/2}|n\rangle .
\end{equation}
As with attenuation and the binomial states, there is no suggestion that this
is the state produced by amplification, but merely that it has the same photon
statistics.  The link with amplification is simply one reason to consider the
properties of the negative binomial states, some of the properties of which are
presented in Appendix \ref{AppendixC}.

The action of the creation operator on the negative binomial state
$|\eta,-(M+1)\rangle$ produces another negative binomial state, but with $M$
increased by unity, as in Eq.~(\ref{EqC11}).  Hence the negative factorial
moments for an $\ell$-photon added negative binomial state are those for a
negative binomial state with $M$ increased by $\ell$:
\begin{eqnarray}
\label{Eq5.10}
\langle (\hat{n}+1)^{(-m)}\rangle_{|\eta,-(M+1)\rangle}^{\ell -} &=&
\langle (\hat{n}+1)^{(-m)}\rangle_{|\eta,-(M+\ell+1)\rangle}  \nonumber \\
&=& \eta^{-m}\frac{(M+\ell+m)!}{(M+\ell)!} .
\end{eqnarray}
We note that, as with the corresponding result for the binomial states, this
expression tells us that the form of the state produced by a combination of
amplification and photon addition does not depend on the order in which these
processes are applied.

\subsubsection{Agarwal's negative binomial states}

As noted above, Agarwal defined negative binomial states somewhat differently,
with a photon number probability distribution starting at $n=0$ rather than at
$n=M$ so that the photon number probability distribution is \cite{AgarwalNeg}
\begin{equation}
\label{Eq5.11}
P_{\rm Agar}(n) = \left(\begin{array}{c}
                                            n+s \\
                                            n \end{array}\right)
\beta^{s+1}(1-\beta)^n .
\end{equation}
To see the connection with the states $|\eta,-(M+1)\rangle$ let us rewrite these
probabilities in a different notation:
\begin{equation}
\label{Eq5.12}
P_{\rm Agar}(n) = \left(\begin{array}{c}
                                            n+M \\
                                            M \end{array}\right)
\eta^{M+1}(1-\eta)^n .
\end{equation}
It is clear from this that 
\begin{equation}
\label{Eq5.13}
P_{\rm Agar}(n) = P_{|\eta,-(M+1)\rangle}(n+M) .
\end{equation}
The moment generating function of the second kind for this state is
\begin{equation}
\label{Eq5.14}
\mathcal{N}_{\rm Agar}(\lambda) = \frac{1}{1+\lambda}\left[\frac{\eta(1+\lambda)}{\lambda+\eta}\right]^{M+1}
\end{equation}
from which it is straightforward to calculate the negative factorial moments.
For the first of these we find
\begin{eqnarray}
\label{Eq5.15}
\langle\hat{n}+1\rangle &=& \left.-\frac{d}{d\lambda} \mathcal{N}_{\rm Agar}(\lambda)\right|_{\lambda = 0} \nonumber \\
&=& \frac{M+1}{\eta} - M .
\end{eqnarray}
We note that this is $M$ less than the corresponding value for the state $|\eta,-(M+1)\rangle$, as it should be.

\section{Conclusions}

Experiments realising both photon subtraction and photon addition have been
shown to lead to novel quantum states~\cite{Philippe} and have been employed to
test one of the most fundamental ideas in quantum optics
\cite{Bellini1,Bellini2,Bellini3}.  It has been shown, moreover, that these
processes can lead to, at first sight, surprising phenomena in optical
measurements~\cite{Bobexpt,Lanz}.  These developments motivated the study
presented here.  We have shown how the statistics of the states produced by
photon subtraction and photon addition can be derived directly and simply from
those of the original state.  The natural tools for this are the moment
generating function $\mathcal{M}(\mu)$, familiar to quantum optics
\cite{Methods}, and a second, closely related function $\mathcal{N}(\lambda)$,
which we introduce here.

We have presented a comprehensive study of the statistics of photon subtracted
and photon added states.  We have found, in particular, that photon subtraction
will result in a increase in the mean photon number if the initial state was
super-Poissonian and that successful photon addition will, except for an
initial number state, increase the mean photon number by more than the number
of photons added and that photon subtraction leaves the mean photon number, and
indeed the full probability distribution, unchanged.  We have seen that the
resolution of these apparently paradoxical behaviours lies in the fact that the
processes are necessarily probabilistic and that the photon number probability
distribution for the incident light \emph{given} that the subsequent process of
subtraction or addition is successful is not the same as the initial
distribution.  The explanation for these behaviours lies, as is so often the
case, in the correct application of Bayes' theorem.

\begin{acknowledgments}

We thank Bob Boyd for rekindling our interest in this topic, Adrian Bowman for
helpful advice on ascending factorials, Tom Douce for assistance with
continuous variable quantum computing and Bill Phillips for suggesting the
example of subtracting a photon from an initial mixture of the vacuum and the
ten photon states.  This work was supported by the UK Engineering and Physical
Sciences Research Council, under the grants EP/R008264/1 and EP/P015719/1, and
by a Royal Society Research Professorship, grant number RP150122.

\end{acknowledgments}

\appendix
\def\normOrd#1{\mathop{:}\nolimits\!#1\!\mathop{:}\nolimits}
\def\antinormOrd#1{\mathop{\scriptscriptstyle{\bm{\vdots}}}\nolimits\!#1\!\mathop{\scriptscriptstyle{\bm{\vdots}}}\nolimits}
\def\antinormOrdtext#1{\mathop{\scriptscriptstyle{\pmb{\bm{\vdots}}}}\nolimits\!#1\!\mathop{\scriptscriptstyle{\pmb{\bm{\vdots}}}}\nolimits}
\section{Moment generating functions}
\label{AppendixA}

\subsection{$\mathcal{M}(\mu)$}

Our first moment generating function is
\begin{equation}
\label{EqA1}
\mathcal{M}(\mu) = \sum_{n=0}^\infty (1-\mu)^n P(n) .
\end{equation}
This function was once a commonly employed tool in quantum optics. Its values
and those of its derivatives provide a wealth of information. In particular the
derivatives evaluated at $\mu = 1$ give the photon-number probabilities:
\begin{equation}
\label{EqA2}
P(n) = \frac{1}{n!}\left. \left(-\frac{d}{d\mu}\right)^n \mathcal{M} \right|_{\mu = 1} .
\end{equation}
The derivatives evaluated at $\mu = 0$ give the factorial moments:
\begin{eqnarray}
\label{EqA3}
\langle \hat n^{(m)} \rangle &=& \langle \hat n(\hat n-1) \cdots (\hat n-m+1)\rangle  \nonumber \\
&=& \left. \left(-\frac{d}{d\mu} \right)^m \mathcal{M}(\mu)\right|_{\mu = 0} .
\end{eqnarray}
The first few of these are
\begin{eqnarray}
\label{EqA4}
\mathcal{M}(0) &=& 1 \nonumber \\
-\frac{d}{d\mu} \mathcal{M}(0) &=& \langle \hat n \rangle  \nonumber \\
\frac{d^2}{d\mu^2} \mathcal{M}(0) &=& \langle \hat n(\hat n-1) \rangle = \langle \normOrd{\hat{n}^2} \rangle ,
\end{eqnarray}
where the dots $\normOrd{~}$ denote normal ordering. More generally the
factorial moment $\langle \hat n^{(m)} \rangle$ is the normal ordered
expectation value of the $\hat n^m$, so that $\langle \hat n^{(m)} \rangle =
\langle \normOrd{\hat n^m} \rangle = \langle  \hat a^{\dagger m} \hat
a^m\rangle $.

We can also extract the moments of the photon number by differentiation. To
this end we introduce the change of variable, $x = \ln(1-\mu)$, so that
\begin{equation}
\label{EqA5}
\mathcal{M} = \sum_{n=0}^\infty e^{nx} P(n) .
\end{equation}
It follows that the required moments are simply derivatives with respect to $x$
evaluated at $x = 0$ (or $\mu = 1$):
\begin{equation}
\label{EqA6}
\langle \hat n^m \rangle = \left. \left(\frac{d}{dx}\right)^m \mathcal{M}\right|_{x = 0} .
\end{equation}

One additional property that we make use of is the fact that $\mathcal{M}(2)$
reveals the probabilities that the number of photons is either even or odd:
\begin{equation}
\label{EqA7}
\mathcal{M}(2) = P({\rm even}) - P({\rm odd}) .
\end{equation}

Part of the utility of the moment generating function arises from the fact that
its evolution can be readily calculated in a number of situations including
both linear amplification and loss.  It is also possible to evaluate it
directly from the quasi-probability phase-space distributions. In particular,
it has a simple form in terms of the Glauber-Sudarshan P function:
\begin{equation}
\label{EqA8}
\mathcal{M}(\mu) = \int d^2\alpha e^{-\mu |\alpha|^2}P(\alpha) .
\end{equation}
This follows directly from the operator ordering theorem
\begin{equation}
\label{EqA9}
(1 - \mu)^{\hat n} =\normOrd{e^{-\mu \hat n}} \,.
\end{equation}
If we take the expectation value of this operator we find an expression for the
moment generating function in terms of the factorial moments:
\begin{equation}
\label{EqA10}
\mathcal{M}(\mu) = \sum_{m=0}^\infty \frac{(-\mu)^m}{m!}\langle \hat{n}^{(m)} \rangle ,
\end{equation}
the Maclaurin series of which gives the factorial moments as in
Eq.~(\ref{EqA3}).  It should be emphasised, however, that the integral form
Eq.~(\ref{EqA8}) may run into convergence problems for some states and for
certain values of $\mu$.  When such difficulties arise, the original form
Eq.~(\ref{EqA1}) should be used.

\subsection{$\mathcal{N}(\lambda)$}

Our second moment generating function is
\begin{equation}
\label{EqA11}
\mathcal{N}(\lambda) = \sum_{n=0}^\infty (1+\lambda)^{-(n+1)} P(n) .
\end{equation}
The first thing that should be noted is that this function is simply related to
the first moment generating function,
\begin{eqnarray}
\label{EqA12}
\mathcal{N}(\lambda) &=& \frac{1}{1+\lambda}\mathcal{M}\left(\frac{\lambda}{1+\lambda}\right) \nonumber \\
\mathcal{M}(\mu) &=& \frac{1}{1-\mu}\mathcal{N}\left(\frac{\mu}{1 - \mu}\right) ,
\end{eqnarray}
but it proves convenient to introduce it as a separate function because of its
distinctive properties.  Principal among these is the ease with which we can
generate negative or ascending factorial moments:
\begin{equation}
\label{EqA13}
\langle (\hat n + 1)^{(-m)}\rangle = \langle (\hat n + 1)(\hat n + 2)\cdots (\hat n + m)\rangle  ,
\end{equation}
where $x^{(-m)}$ denotes the ascending factorial~\cite{Steffensen} or
Pochhammer symbol~\cite{Abram,Olver}
\begin{equation}
\label{EqA14}
x^{(-m)} = x(x+1)\cdots(x + m -1)
\end{equation}
The negative factorial moments are simply the expectation values of the
corresponding powers of the number operator in antinormal order:
\begin{equation}
\label{EqA15}
\langle (\hat n + 1)^{(-m)}\rangle = \langle \antinormOrd{\hat n^m}\rangle ,
\end{equation}
where the dots $\antinormOrdtext{~}$ denote antinormal order, that is
$\langle\antinormOrd{\hat n^m}\rangle = \langle   \hat a^m \hat a^{\dagger
m}\rangle$. The negative factorial moments are obtained from $\mathcal{N}(\mu)$
by differentiation in an analogous manner to the factorial moments from
$\mathcal{M}(\mu)$:
\begin{equation}
\label{EqA16}
\langle (\hat n + 1)^{(-m)}\rangle = \left. \left(-\frac{d}{d\lambda} \right)^m \mathcal{N}(\lambda)\right|_{\lambda = 0} .
\end{equation}

We note also that the function $\mathcal{N}(\lambda)$ provides other
information including the probability that the photon number is even or odd:
\begin{equation}
\label{EqA17}
\mathcal{N}(-2) = P({\rm odd}) - P({\rm even}) = -\mathcal{M}(2) .
\end{equation}
It is also simply related to the Husimi or Q quasi-probability distribution:
\begin{equation}
\label{EqA18}
\mathcal{N}(\lambda) = \int d^2\alpha e^{-\lambda|\alpha|^2} Q(\alpha) ,
\end{equation}
which follows from the operator identity
\begin{equation}
\label{EqA19}
(1+\lambda)^{-(\hat n + 1)} =\antinormOrd{e^{-\lambda \hat n}} \, .
\end{equation}
If we take the expectation value of this operator we find an expression for the moment generating
function in terms of the negative factorial moments:
\begin{equation}
\label{EqA20}
\mathcal{N}(\lambda) = \sum_{m=0}^\infty \frac{(-\lambda)^m}{m!}\langle (\hat{n}+1)^{(-m)} \rangle ,
\end{equation}
the Maclaurin series of which gives the negative factorial moments as in
Eq.~(\ref{EqA16}). As with our first moment generating function, the integral
form given here, in Eq.~(\ref{EqA18}), may have convergence problems for some
values of $\lambda$.  In such cases the original form, Eq.~(\ref{EqA11}) should
be used.

Finally, we note that the photon number probability distribution can be obtained from $\mathcal{N}(\lambda)$
by differentiation:
\begin{equation}
\label{EqA21}
P(n) = \lim_{\lambda \rightarrow \infty} \frac{(1+\lambda)^{n+1}}{n!}
\left(-\frac{d}{d\lambda}\right)^n (1+\lambda)^n \mathcal{N}(\lambda) .
\end{equation}

\section{Derivation of an inequality}
\label{AppendixB}

We require the inequality
\begin{equation}
\label{EqB1}
\langle (\hat{n}+1)(\hat{n}+1)\cdots (\hat{n} + \ell)\rangle \geq \langle(\hat{n}+1)\cdots (\hat{n} + \ell)\rangle
\langle (\hat{n}+1)\rangle
\end{equation}
in the derivation of Eq.~(\ref{Eq4.15}).  To establish this let us consider,
first, a more general combination:
\begin{eqnarray}
\label{EqB2}
\langle B(\hat{n})A(\hat{n})\rangle &-& \langle B(\hat{n})\rangle\langle A(\hat{n})\rangle  \nonumber \\
\qquad = & &   \sum_n P(n) B(n)A(n) \nonumber \\
& & \quad - \sum_n\sum_m P(n)P(m)B(n)A(m)  \nonumber \\
\qquad = & &   \sum_n \sum_m P(n)P(m) B(n)A(n) \nonumber \\
& & \quad - \sum_n\sum_m P(n)P(m)B(n)A(m)  \nonumber \\
\qquad = & & \frac{1}{2} \sum_m\sum_n P(n)P(m) [A(m) - A(n)]  \nonumber \\
& & \qquad \times[B(m) - B(n)] .
\end{eqnarray}
For $A(n) = n+1$ and $B(n) = (n+1)\cdots(n+\ell)$ the combinations $A(m) -
A(n)$ and $B(m) - B(n)$ are either both positive or both negative for all $m
\neq n$ and hence the terms in the summation are all greater than or equal to
zero and the inequality Eq.~(\ref{EqB1}) follows.  Note that for this reason
the equality in Eq.~(\ref{EqB1}) holds if and only if $P(n) = \delta_{n,N}$ for
some $N$ corresponding to photon number state.

\section{Binomial and negative binomial states}
\label{AppendixC}

The binomial and negative binomial states are pure states for which the photon
number probabilities correspond to the binomial and negative binomial
distributions respectively.  We summarise here some of the more important
properties of these states.

\subsection{Binomial states}

The binomial states are defined to be pure states with a photon number
probability distribution that is of binomial form~\cite{Stoler}:
\begin{equation}
\label{EqC1}
|\eta,M\rangle = \sum_{n=0}^M\beta_n^M  |n\rangle ,
\end{equation}
where
\begin{equation}
\label{EqC2}
\beta_n^M = \left[\left(\begin{array}{c}
                             M \\
                             n \end{array}\right)\eta^n(1-\eta)^{M-n}\right]^{1/2} .
\end{equation}
Here $M$ is a non-negative integer and $\eta$ can take any value between $0$ and $1$.  The
action of the annihilation operator on this state produces another binomial state, one with $M$
reduced by unity:
\begin{equation}
\label{EqC3}
\hat{a}|\eta,M\rangle = \sqrt{\eta M}|\eta,M-1\rangle .
\end{equation}
It follows that the mean photon number for this state is $\eta M$ and, more generally, the factorial moments
for this state are
\begin{equation}
\label{EqC4}
\langle \hat{n}^{(m)}\rangle = \eta^m \frac{M!}{(M-m)!} .
\end{equation}
This means, in particular, that the states exhibit sub-Poissionian statistics, with a normally-ordered
photon number variance that is negative:
\begin{equation}
\label{EqC5}
:\Delta n^2: = \langle \hat{n}^{(2)}\rangle - \langle \hat{n}\rangle^2 = -\eta^2M .
\end{equation}

If we generalise the states to include a phase,
\begin{equation}
\label{EqC6}
|\eta,M,\theta\rangle = \sum_{n=0}^M\beta_n^M  e^{in\theta} |n\rangle ,
\end{equation}
then we have an over-complete set of states.  To see this we need only note that the states are, in general,
not orthogonal but can be used to represent the identity operator:
\begin{eqnarray}
\label{EqC7}
\frac{1}{2\pi}\int_0^{2\pi} d\theta & \eta & \sum_{M=0}^\infty   |\eta,M,\theta\rangle \langle\eta,M,\theta| \nonumber \\
& & \qquad  = \sum_{M=0}^\infty \sum_{n=0}^M \eta \left(\beta_n^M \right)^2 |n\rangle \langle n|  \nonumber \\
& & \qquad  = \sum_{n=0}^\infty |n\rangle \langle n| \nonumber \\
& & \qquad  = \hat{\rm I} .
\end{eqnarray}
For example, the mixed state produced by attenuating the photon number state $|M\rangle$ has the density operator
\begin{equation}
\label{EqC8}
\hat{\rho}^{\rm Att.}_{|M\rangle} =
\frac{1}{2\pi}\int_0^{2\pi} d\theta   |\eta,M,\theta\rangle \langle\eta,M,\theta| .
\end{equation}
Further properties of this state may be found in~\cite{Stoler}.

\subsection{Negative binomial states}

The negative binomial states are defined to be pure states with a photon number
probability distribution that is of negative binomial form~\cite{negbinomial}:
\begin{equation}
\label{EqC9}
|\eta, -(M+1)\rangle = \sum_{n=M}^\infty \beta_n^{-(M+1)}|n\rangle ,
\end{equation}
where
\begin{equation}
\label{EqC10}
\beta_n^{-(M+1)}
 = \left[\left(\begin{array}{c}
                             n \\
                             M \end{array}\right)\eta^{M+1}(1-\eta)^{n-M}\right]^{1/2} .
\end{equation}
Here $M$ is again a non-negative integer and $\eta$ can take any value between $0$ and $1$.  For these states
it is the action of the creation operator that is simple:
\begin{equation}
\label{EqC11}
\hat{a}^\dagger|\eta, -(M+1)\rangle = \sqrt{\frac{M+1}{\eta}}|\eta, -(M+2)\rangle .
\end{equation}
It follows that the mean photon number for this state is $(M+1)/\eta - 1$ and, more generally, that the negative
factorial moments for this state have the form
\begin{equation}
\label{EqC12}
\langle (\hat{n}+1)^{(-m)}\rangle = \eta^{-m}\frac{(M+m)!}{M!} ,
\end{equation}
so that the antinormally ordered variance in the photon number is
\begin{equation}
\label{EqC13}
\dot{:}\Delta n^2\dot{:} = \langle (\hat{n}+1)^{(-2)}\rangle - \langle (\hat{n}+1)\rangle^2
= \frac{M+1}{\eta^2} .
\end{equation}

As with the binomial states, we can generalise the negative binomial states by including a phase
\begin{equation}
\label{EqC14}
|\eta,-(M+1),\theta\rangle = \sum_{n=0}^M\beta_n^{-(M+1)}  e^{in\theta} |n\rangle ,
\end{equation}
with the resulting set of states being overcomplete so that they form a resolution of the identity:
\begin{eqnarray}
\label{EqC15}
\frac{1}{2\pi}\int_0^{2\pi} d\theta & \eta^{-1} & \sum_{M=0}^\infty   |\eta,-(M+1),\theta\rangle \langle\eta,-(M+1),\theta| \nonumber \\
& & \qquad = \sum_{M=0}^\infty \sum_{n=M}^\infty \eta^{-1} \left(\beta_n^{-(M+1)} \right)^2 |n\rangle \langle n|  \nonumber \\
& & \qquad  = \sum_{n=0}^\infty |n\rangle \langle n| \nonumber \\
& & \qquad  = \hat{\rm I} .
\end{eqnarray}
In particular, the mixed state produced by amplifying the photon number state $|M\rangle$ with a gain $G = \eta^{-1}$
has the density operator
\begin{equation}
\hat{\rho}^{\rm Amp.}_{|M\rangle} =
\frac{1}{2\pi}\int_0^{2\pi} d\theta   |\eta,-(M+1),\theta\rangle \langle\eta,-(M+1),\theta| .
\end{equation}
Further properties of this state may be found in~\cite{negbinomial}.

\end{document}